\documentclass[prb,aps,twocolumn]{revtex4}
\usepackage{epsfig}

\begin{document}

\title{Instability of the rhodium magnetic moment as origin 
of the metamagnetic phase transition in $\alpha$-FeRh}
\author{M. E. Gruner}\email{me@thp.uni-duisburg.de}
\author{E. Hoffmann}
\author{P. Entel}
\affiliation{Theoretische Tieftemperaturphysik,
Gerhard-Mercator-Universit\"at Duisburg, 47048 Duisburg, Germany}
\date{\today}
\begin{abstract}
Based on {\em ab initio} total energy calculations
we show that two magnetic states of
rhodium atoms together with competing ferromagnetic and
antiferromagnetic exchange interactions are responsible
for a temperature induced
metamagnetic phase transition, which experimentally is observed
for stoichiometric $\alpha$-FeRh.
Taking into account the results of previous and newly performed
first-principles calculations we present a spin-based model, which
allows to reproduce this first-order metamagnetic transition
by means of Monte Carlo simulations.
Further inclusion of spacial variation of exchange parameters
leads to a realistic description of the experimental magneto-volume
effects in $\alpha$-FeRh.
\end{abstract}

\pacs{75.30.Kz, 75.10.Hk, 71.15.Nc, 75.50.Bb}

\maketitle

\section{Introduction}
\begin{figure}
  \begin{center}
    \epsfig{file=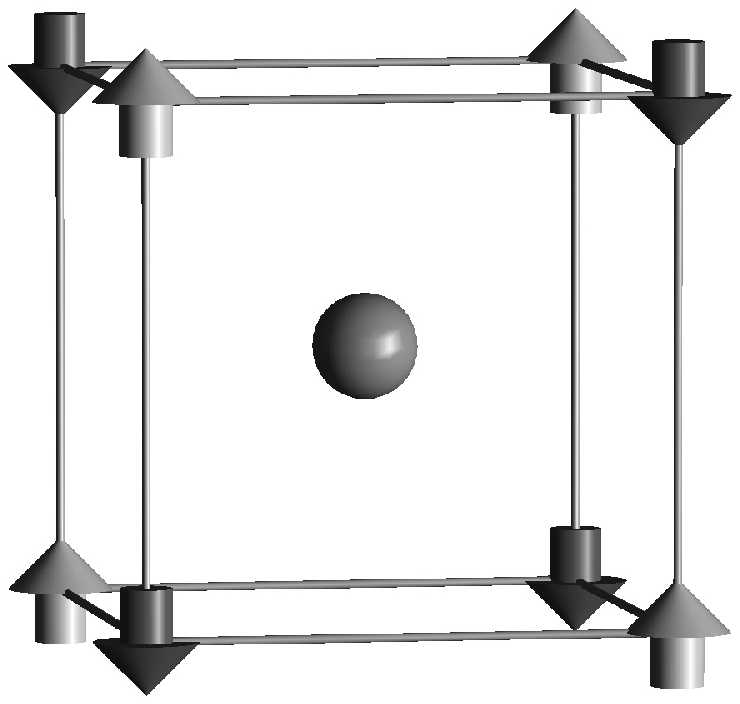, width=1.4566667in}\hfill
    \epsfig{file=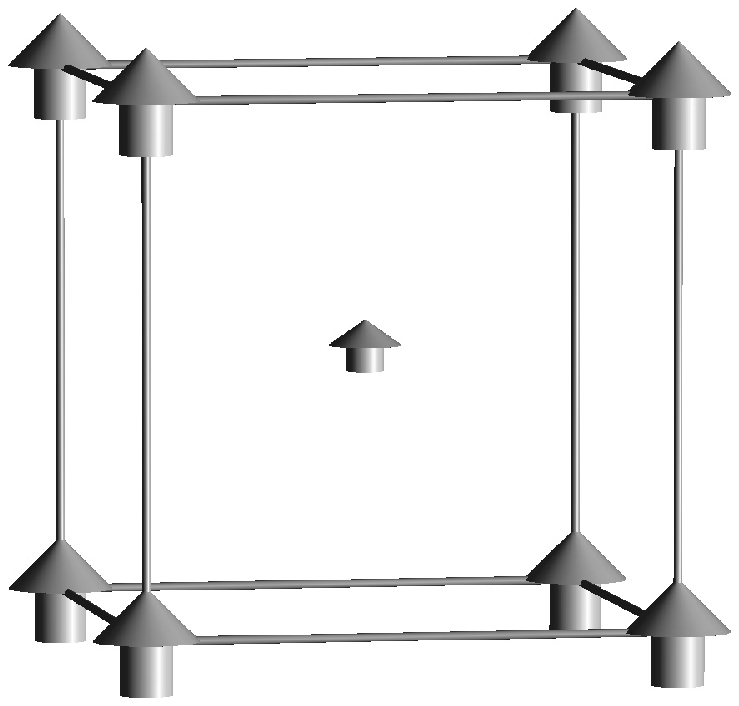, width=1.4566667in}
  \end{center}  
  \caption{Left: Type II antiferromagnetic ground state structure with
    a nonmagnetic Rh atom at the center and Fe atoms with moments
    of $3.3\,\mu_{\rm B}$ at the corner site. Right: Ferromagnetic
    structure with Fe moments of $3.2\,\mu_{\rm B}$ at the corners and
    Rh moment of $0.9\,\mu_{\rm B}$ at the center site.\label{fig:structure}
    }
\end{figure}

In 1938 Fallot\cite{cn:Fallot38,cn:Fallot39} discovered that
ordered bcc Fe$_{50}$Rh$_{50}$
undergoes a first order metamagnetic transition from an antiferromagnetic (AF)
ground state to a ferromagnetic (FM) phase with increasing temperature.
This transition occurs at
$T_{\rm m}\approx 320\,{\rm K}$ and is
accompanied by a volume increase of
about $1\,\%$.\cite{cn:Kouvel66,cn:Zakharov64TR}
The Curie temperature $T_{\rm C}$ of the FM phase is of the order of
$670\,{\rm K}$.\cite{cn:Zakharov64TR,cn:Kouvel62}
In contradiction to the first hypothesis of Fallot and
Horcart,\cite{cn:Fallot39}
X-ray diffraction measurements showed that the transition 
is isostructural.\cite{cn:Bergevin61} 
From M\"ossbauer and neutron diffraction
measurements one knows that the FM phase has
collinear magnetic moments of $3.2\,\mu_{\rm B}$ per Fe atom and
$0.9\,\mu_{\rm B}$ per Rh atom.\cite{cn:Shirane63} At low temperatures
an AF-II spin structure is found with Fe moments of
$3.3\,\mu_{\rm B}$ and with vanishing
Rh moments (see Figure~\ref{fig:structure}).\cite{cn:Shirane64,cn:Kunitomi71}
Application of hydrostatic pressure suppresses the FM
phase, i.\,e. for a critical pressure of $60\,{\rm kbar}$ 
the system immediately
transforms from the AF to the paramagnetic (PM) phase.

An early explanation for this behavior was based on
the phenomenological {\em exchange inversion
model\/}\cite{cn:Kittel60}
by Kittel (which originally was designed to explain metamagnetic
transitions in other materials
like Cr-modified Mn$_2$Sb with 
a layered magnetic structure).
In this model the exchange
parameter varies linearly with the lattice constant and changes
sign for a critical value $a_{\rm c}$. However,
experimental findings like the rather large entropy change%
\cite{cn:Kouvel66,cn:Zakharov64TR,cn:Annaorazov96,cn:Lommel69,cn:McKinnon70}
at $T_{\rm m}$, which is of the order of 
$\Delta S(T_{\rm m})\approx 12.5-19.7\,
{\rm J\,kg^{-1}K^{-1}}$, as well as elastic properties 
could not correctly be described.\cite{cn:McKinnon70}
Tu {\it et al.}\cite{cn:Tu69} used a different approach by considering
the large difference of the low-temperature specific heat 
constants, $\gamma$, of
the AF and the FM phases: Measurements suggest that
$\gamma_{\rm FM}\approx 59-62.5\,{\rm mJ\,kg^{-1}K^{-2}}$ is about four
to six times
larger than $\gamma_{\rm AF}\approx 10.5-16\,
{\rm mJ\,kg^{-1}K^{-2}}$.\cite{cn:Tu69,cn:Ivarsson71}
Based on these observations, Tu {\it et al.} explained the transition 
by a change in
entropy of the band electrons between the AF and the FM phases. 
An estimation of the free energy
shows that these contributions have the right order of magnitude to explain the
AF--FM transition, if one assumes that the electronic densities of states 
at the Fermi level do not vary considerably 
from low temperatures up to the transition.
However, since the FM phase cannot be stabilized at low temperatures for
stoichiometric FeRh, $\gamma_{\rm FM}$ has been measured in slightly more
iron-rich alloys. With respect to the strong sensitivity of the
magnetic phase diagram to small departures
from stoichiometry,\cite{cn:Vinokurova88b}
one may doubt
that the low temperature specific heat can be considered to be
independent of concentration.\cite{cn:Ponomarev73TR,cn:Moruzzi92}
A further drawback considering the explanation of
Tu {\it et al.}\/\ arises from the fact that adding $5\,\%$ Iridium boosts
$\gamma_{\rm AF}$ by almost an order of magnitude to a value of 
$\gamma_{\rm AF}=101\,{\rm mJ\,kg^{-1}K^{-2}}$:%
\cite{cn:Ivarsson71,cn:Fogarassy72}
The compound Fe$_{49.5}$Rh$_{45.5}$Ir$_{5}$ also undergoes a 
metamagnetic transition with $T_{\rm m}$ shifted to higher
temperatures.\cite{cn:Kouvel66}
Since the relation between $\gamma_{\rm AF}$ and
$\gamma_{\rm FM}$ is reversed in this material, the previously sketched 
explanation cannot be applied to this case.

In 1992 Moruzzi
and Marcus performed  {\em ab initio} calculations
using spin polarized density functional theory (DFT) in the
framework of the local density
approximation (LDA) and the augmented
spherical wave (ASW) formalism.\cite{cn:Moruzzi92,cn:Moruzzi92b,cn:Moruzzi93}
They calculated the total energy of FeRh as a function of
volume for different magnetic structures. They found that the AF-II
spin structure is the ground state, whereas the FM structure is
another stable solution with higher energy and a larger volume.
For the moments they obtained
$\mu_{\rm Fe}=2.98\,\mu_{\rm B}$,
$\mu_{\rm Rh}=1.02\,\mu_{\rm B}$ (AF phase) and
$\mu_{\rm Fe}=3.15\,\mu_{\rm B}$
(FM phase), in agreement with
experimental results.

In this work, we present further {\em ab initio\/} total energy
calculations of stoichiometric $\alpha$-FeRh using the ASW
formalism and the generalized gradient
approximation (GGA).\cite{Perdew:92a} The corresponding code also allows 
to evaluate noncollinear alignment
of spins (see Ref.\ \onlinecite{Uhl:94} and
references therein).
For many cases LDA gives a reasonable description of the
ground state properties of solids. However, in some cases it
predicts the wrong ground state as e.\,g.\ for
iron\cite{Stixrude:94}. For Fe,
the hcp structure is found to be $10\,$mRy lower in energy than the 
bcc phase.
Because FeRh has also a bcc ground state lattice structure (unlike pure Rh),
it is necessary to go beyond LDA by using GGA.
We have considered different possible magnetic ground state
structures presented in
section~\ref{sec:abinitio}.
Based then on the specific energetic order of resulting
energy versus volume curves, we discuss a new mechanism
for the temperature driven metamagnetic transition in FeRh.
In contrast to prior explanations, our model does not rely
on ground state properties or low temperature data alone.
Instead, we propose that thermal excitations at finite temperatures
are the driving force for the transition.
This is demonstrated in section~\ref{sec:simple} on the basis of
Monte Carlo simulations
of an Ising-type spin model, showing that a competition
between AF Fe-Fe exchange-interactions and a nonmagnetic Rh state on
the one hand and FM Fe-Rh interactions on the other
is sufficient to explain the metamagnetic transition of FeRh.
Inclusion of spatial degrees of freedom (section~\ref{sec:extended}) 
by adding pair potentials and assuming a linear variation of the
exchange  parameter with the lattice parameter
proves that this model explanation is in
accordance with the {\em ab initio} data and leads furthermore to
a nearly quantitative description of experimental details.

\section{Ab initio total energy calculations}\label{sec:abinitio}
\begin{table}
\begin{ruledtabular}\begin{tabular}{lccccccccc}
\multicolumn{2}{c}{$V_{\rm 0}$}& $\Delta E$&
\multicolumn{2}{c}{$m_{\rm Fe}$} & \multicolumn{2}{c}{$B$} &
\multicolumn{2}{c}{$N(E_{\rm F})$} \\
& \multicolumn{2}{c}{a.u./atom} &
mRy & \multicolumn{2}{c}{$\mu_{\rm B}$} & \multicolumn{2}{c}{GPa} &
\multicolumn{2}{c}{states/Ry} \\
&AF&FM&&AF&FM&AF&FM&AF&FM \\ \hline
Ref. \onlinecite{cn:Moruzzi92} & 90.2 & 91.8 & 1.9 & 2.98 & 3.15 &
214 & 202 & 37 & 32 \\
Ref. \onlinecite{cn:Szajek94} & 91.9 & 92.9 & 2.2 & 3.13 & 3.20 &
227 & 244 & 13 & 32 \\
This work& 91.4& 93.0 & 2.5 &3.18 &3.23 & 197 & 193 & 13 & 29 \\
\end{tabular}\end{ruledtabular}
  \caption{Comparison of ab initio results.Here,  $V_{\rm 0}$ is the
    equilibrium volume of
    AF and FM phases, $\Delta E=E_{\rm
    FM}-E_{\rm AF}$ is the energy difference per atom, 
    $m_{\rm Fe}$ is the Fe magnetic moment, $B$ is the bulk modulus, and
    $N(E_{\rm F})$ the density of states per formula unit at the Fermi level. 
    The moment on the Rh sites
    is $1.0\,\mu_{\rm B}$ in the FM phase and zero in the AF phase in
    all calculations.}
    \label{tab:ab_initio}
\end{table}

We have performed {\em ab initio} calculations by using 
the ASW-method \cite{Willams:79} and GGA.
Relativistic effects are included in the scalar relativistic approximation.
The basis wave functions of Fe and Rh atoms include {\em spd\/} and {\em f\/}
states which are sufficient to obtain the correct
magnetic behavior of $\gamma$-iron.
We assume that the AF like spin structure can best be described by
a spin-spiral (Ref. \onlinecite{Uhl:94}) with wave vector 
$(0.5,0.5,0.5)$ in units of $2\pi/a$.
As first step, we optimized the volume of the AF-II state with 
equal ASA-radii for both types of atoms.
For the resulting equilibrium volume, the total energy
is then minimized with respect to 
the ratio of the ASA-spheres leading to
$r_{\rm Rh}/r_{\rm Fe}=1.15$. 
In order to investigate the influence of the Rh moment in the
FM phase we have also used the fixed spin moment
method \cite{Uhl:94}
to restrict $m_{\rm Rh}$ to zero.
\begin{figure}
  \begin{center}  
    \epsfig{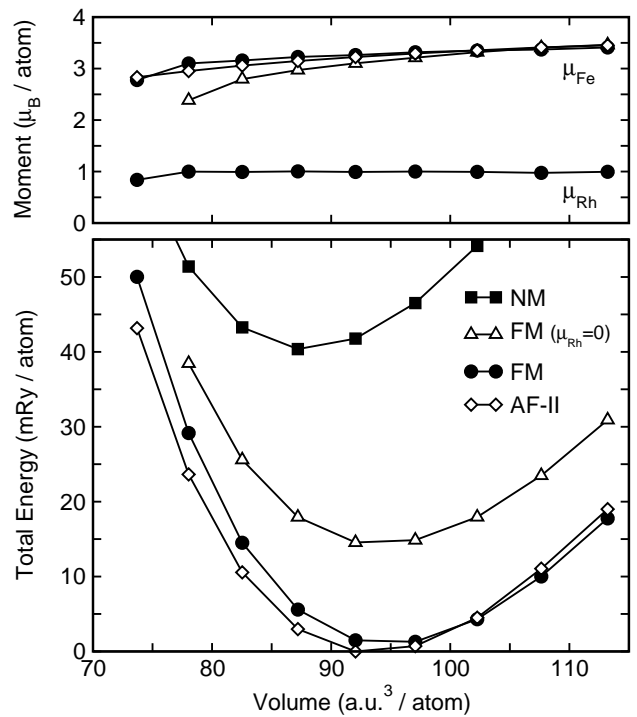}\hfill
  \end{center}  
  \caption{Total energy and magnetic moment versus atomic volume as
          obtained by {\em ab initio} calculations
          (present work).\label{fig:ASW}}
\end{figure}
Table \ref{tab:ab_initio} contains the calculated
equilibrium properties for AF-II and FM phases which
are in fair agreement with previous calculations. 
Compared to previous non-relativistic ASW LDA calculations
(Moruzzi {\it et al.}\cite{cn:Moruzzi92}),
we obtain equilibrium volumes which are about 1-2$\,\%$ larger, being
typical for GGA. 
In contrast to the cases of pure iron\cite{cn:Heike99} and $\mathrm{FeNi}$ 
Invar alloys\cite{Barbiellini:91}, the magnetic
energy differences are only slightly influenced by the gradient corrections.
Also, no evidence was found that a noncollinear structure could be lower in
energy than the previously found collinear AF and FM spin arrangements.
With respect to the density of states at the Fermi level
our results compare better with results of Szajek 
{\it et al.}\cite{cn:Szajek94}
than with results of Ref.~\onlinecite{cn:Moruzzi92}.
Since the latter results were obtained by using LDA ASW
(with a {\em spd\/} basis), we conclude that
the density of states at the Fermi level is very sensitive to
computational details. The values obtained for the bulk moduli, 
$B_{\rm AF}$ and $B_{\rm FM}$, are lower compared to those of
Ref. \onlinecite{cn:Moruzzi92} (where
$B_{\rm AF}>B_{\rm FM}$ was obtained in contrast to
$B_{\rm FM}>B_{\rm AF}$ in Ref.\ \onlinecite{cn:Szajek94}).

The calculated total energy curves are shown in Fig.~\ref{fig:ASW}.
The large energetic difference between the usual FM phase and
the hypothetic
FM phase with zero rhodium moment implies that a finite
Rh moment plays an important role for the stability of the FM phase.
It seems unlikely that a magnetic field at the Rh sites
induced by the surrounding iron atoms is responsible
for the appearance of a Rh moment in the FM phase, as was
proposed in previous discussions.
However, a strong ferromagnetic exchange interaction between the 
Rh and the iron atoms that overrides an antiferromagnetic
exchange between next nearest neighbor iron sites, 
would explain how the existence of a
Rh moment can help to stabilize the FM phase.
We have evidence as shown in the following sections,
that a competition between a low lying nonmagnetic Rh state 
and another one with higher energy and finite moment (which can
benefit from exchange with ferromagnetic iron neighbors) is the
main reason for the metamagnetic transition. 

\section{A spin-analogous model}\label{sec:simple}
In order to confirm our hypothesis we have constructed a model being
suitable for an examination of the metamagnetic transition
at finite temperatures. The simplest way to do this is by means of
Monte Carlo simulations with a localized spin model.
To keep the model tractable, we
neglect spin wave excitations and restrict ourselves to Ising
spins. This is justified because
the transition takes place between two ordered structures and
both phases have collinear spin structures.
For a description 
of the nonmagnetic Rh state in addition to the magnetic Fe and Rh
states, we chose a spin-1 Ising
model, where the spin variables can take the values $S_i=-1,0,+1$.
The spins are located on a bcc lattice with nearest and next nearest neighbor
interactions. Depending on their positions, we
distinguish between Fe or Rh sites, where each type occupies a simple
cubic sublattice, corresponding to an ordered equiatomic
alloy. The interaction parameters depend then on the type of
sites involved.
This situation can be described by the
following Hamiltonian,
\begin{equation}
    H  \!=\!  -\!\sum_{i}^{~}\:D_i\:S_i^2 -
    \!\!\sum_{<{\rm nn,nnn}>}\!\!J_{ik}\:S_i\:S_k\:.\label{eq:simple}
\end{equation}
Without the assumption of different types of atoms, Hamiltonian
(\ref{eq:simple}) is also known as the Blume-Capel
model.\cite{cn:Blume66,cn:Capel66}
The first term separates the nonmagnetic
$S_i=0$ and the magnetic $S_i=\pm 1$ states.
For Fe we choose a
large positive value in order to suppress the $S_i=0$ state. For
Rh we choose a negative value leading to a nonmagnetic ground state.
The second term contains the exchange parameters $J_{ik}$ which
depend only on the type of atoms located at the sites $i$ and $k$.
In the case of ordered equiatomic FeRh, we have only three different
parameters: $J^{\rm nnn}_{\rm FeFe}$, $J^{\rm nn}_{\rm FeRh}$ and
$J^{\rm nnn}_{\rm RhRh}$. The first one is chosen to be negative in
order to accomplish an AF ground state.
The second is taken to be large and positive as
outlined in the previous section. The third one, for the sake of
simplicity, is set equal to zero.
The choice for $J^{\rm nnn}_{\rm FeFe}$ is fixed by
the N\'eel-temperature $T_{\rm N}$ of the AF phase assuming that no
transition to a FM phase takes place. $T_{\rm N}$ can be
determined from the P-T phase diagram by extrapolating the
transition line between AF and PM phases (occurring
at pressures larger than $6\,{\rm GPa}$) to zero pressure.
$J^{\rm nn}_{\rm FeRh}$ and $D_{\rm Rh}$ have been chosen to yield
realistic values for $T_{\rm C}$ and $T_{\rm m}$, respectively.
The values of parameters used in the simulations are given in
Table~\ref{tab:param_simple}.
 
\begin{table}
  \begin{center}
    \begin{ruledtabular}\begin{tabular}{rrrrr}
      $D_{\rm Fe}$ &
      $D_{\rm Rh}$ &
      $J^{\rm nn}_{\rm FeRh}$ &
      $J^{\rm nnn}_{\rm FeFe}$ & 
      $J^{\rm nnn}_{\rm RhRh}$ \\ \hline
      $\gg\!k_{\rm B}\,T$~ & $-11.1$~ &
      $2.13$~ & $-1.00$~ & 
      $0$~ \\
    \end{tabular}\end{ruledtabular}
  \end{center}
  \caption{Parameters for the spin-analogous model
      (in mRy).\label{tab:param_simple}}
\end{table}

\subsection{Details of computation}
The evaluation of thermodynamic properties of
(\ref{eq:simple}) is done on the basis of Monte Carlo simulations
according to the Metropolis scheme\cite{cn:Metropolis53} using a
sequential update.
Interesting quantities like magnetization or magnetic moment are
computed and summed up every $10$ to $20$ lattice sweeps, which
ensures that the evaluated lattice configurations are sufficiently
uncorrelated. Furthermore, we discard the first
$20\,000$ lattice sweeps in order to allow the system to reach thermal
equilibrium before computing averages. In order to speed this up, 
we have also
used the final configuration of the last run to initialize the
simulation for the next temperature. Simulations which involve a phase
transition are performed twice, with increasing and decreasing 
temperatures, in order to assure that thermal equilibrium has been reached.

The computed AF and FM phases are metastable, i.\,e.\ they are separated
by a large energy barrier, which arises from the fact that in the
transition states a considerable amount of FM domains have to be
created in the AF phase and vice versa.
So the standard algorithm
is unlikely to overcome this barrier and as a result
the metamagnetic transition might not be seen at all.
Instead the phases
have to be overheated or undercooled before transforming, which
results in a large hysteresis or irreversible behavior which makes it
difficult to obtain reliable information about the transition point.
Therefore, it is necessary to modify the algorithm in order to allow for a
direct jump to the other phase by circumventing the
energy barrier with a global update step,
where all spins will be updated at once.
This algorithm ideally connects equilibrium
configurations of the AF phase directly
with equilibrium configurations of the FM phase while
ensuring that the entropy difference 
between the states is correctly reproduced.
This can be done by choosing a unique mapping between
lattice configurations of the AF and FM phases, respectively.
Or, more general, the selection probability of a specific target
configuration must be the same as the selection probability
of the previous start configuration in the backward direction.

Since it is a priori not clear how equilibrium configurations will
look like at finite temperatures,
we simply use an update scheme which
connects the ground state configurations of both phases and thus
works at least at low temperatures. 
In the vicinity of the transition temperature, this algorithm might not
reproduce equilibrium states for the trial configuration,
because nature and amount of excitations are presumably different 
in both phases. 
Since off-equilibrium  means that most of the trial states are
too high in energy, the probability that the trial state is by chance
close enough to an equilibrium state to be accepted, decreases
for larger system sizes. 

In order to obtain a trial
configuration, we divide the system into Rh and Fe sublattices.
The Fe sublattice is again divided into two sublattices according to
spin up and spin down positions (see Fig.~\ref{fig:structure}).
For each of these sublattices, it is decided randomly (with probability
$1/2$), whether the corresponding sublattice is flipped as a whole.
For the Rh sublattice another random number decides whether the
spin values $S_i\in\left\{-1,0,+1\right\}$ are rotated
clockwise or counter-clockwise
(i.\,e.\ $-1$ becomes $0$, $0$ becomes $+1$, $+1$
becomes $-1$, or vice versa).
So, each trial configuration is chosen with the same probability.
Afterwards the energy difference $\Delta H$ between the present and
the previous
configuration is computed and the new configuration is accepted with
probability $\min(\exp(-\Delta H/k_{\rm B}T),1)$
assuring {\em detailed balance\/}.
This global update step is performed after each complete lattice sweep.

We performed for each temperature between $100\,000$
and $1\,000\,000$ (around $T_{\rm m}$) lattice sweeps for different linear
system sizes $L=6$ to $16$. However, our global update scheme does
only show a metamagnetic transition within reasonable simulation times
up to a system size of $L=10$. Looking for a different way to
determine the transition point which would also work for larger system
sizes, we estimate the free energy by integrating the specific heat,
\begin{equation}
 F(T) \, = \, E(T) \, - \, T \, \left( \, S(0) \, + \,
   \int_0^T \frac{C}{T}\,\, dT \right)\:. \label{eq:F}
\end{equation}
In the computation we fit the simulated results for the 
specific heat divided by temperature, $C/T$, and the internal energy $E$
with 10th order polynomials,
which can be integrated analytically. For the system
characterized by the Hamiltonian~(\ref{eq:simple}), we neglect the 
entropy contribution at zero temperature $S(0)$, since the ground state spin
structure in both phases is nondegenerate except for systems with spin
inversion symmetry.
Furthermore, concerning the estimation of the free energy finite size
effects are not expected to affect the results, since phase
transitions are not encountered during these simulations. 
For the rest of the calculations a comparison of the results for
smaller and larger systems sizes (as can be seen in the upcoming
figures) reveals that the main issues of this paper are also
not affected by the restricted system sizes.

\subsection{Computational results}
\begin{figure}
  \begin{center}  
    \epsfig{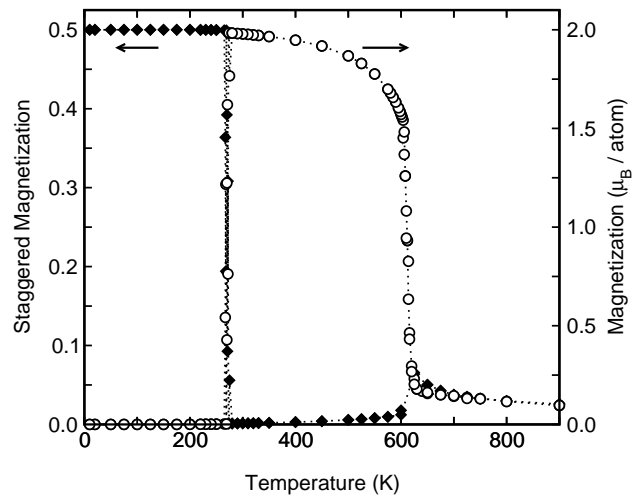}\hfill
  \end{center}  
  \caption{AF order parameter (diamonds) and magnetization (circles) 
    as a function
    of temperature. The magnetization is obtained by multiplying the
    spin values with a value of $3.0\,\mu_{\rm B}$ for Fe and
    $1.0\,\mu_{\rm B}$ for Rh before averaging.
    The simulated system size is $L=10$.\label{fig:M}
}
\end{figure}
\begin{figure}
  \begin{center}  
    \epsfig{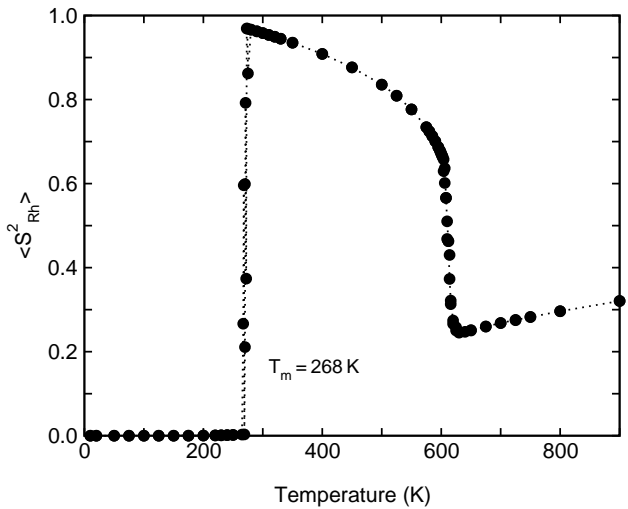}\hfill
  \end{center}  
  \caption{Mean magnetic moment of the Rh atoms.\label{fig:S2}
}
\end{figure}
The order parameter of the AF phase is the staggered magnetization for
the AF-II spin structure, i.\,e.\ the sum (of the
absolute values) of the staggered magnetizations of the two simple cubic
sublattices that constitute the bcc lattice structure.
The staggered magnetization of a simple cubic (NaCl) lattice is defined
as sum of spins multiplied with a sign which alternates
depending on whether the corresponding spin occupies a Na or a Cl position.
The order parameter of the FM phase is given
by the magnetization of the lattice.
The variation of both order parameters with temperature is shown in
Fig.~\ref{fig:M} for a system of size $L=10$.
At low temperatures the staggered magnetization approaches a maximum value of
$0.5$ due to the fact that the Rh sublattice has no moment
(Fig.~\ref{fig:S2}) and therefore does not contribute to the sum.
At $T_{\rm m}=268\,$K the staggered magnetization abruptly drops to
zero, whereas
magnetic moment and magnetization increase close to their
saturation values. Above $T_{\rm C}=610\,$K ferromagnetism breaks down and
the system becomes paramagnetic. At the same time, the average moment of the
Rh atoms falls down to a value of $0.2\,\mu_{\rm B}$.
This is also the reason, why the decrease in the magnetization appears
unusually sharp. From the data, a phase transition of second as well
as of first order seems to be possible.
Accordingly, it is known that for negative values of
$D$ the Blume-Capel model can show both kinds of phase
transitions separated by a tricritical point in the
$T$-$D$-plane.\cite{cn:Blume71,cn:Saul74,cn:Jain80}
Hence in order to safely determine the nature of
this phase transition further calculations are necessary.

\begin{figure}
  \begin{center}  
    \epsfig{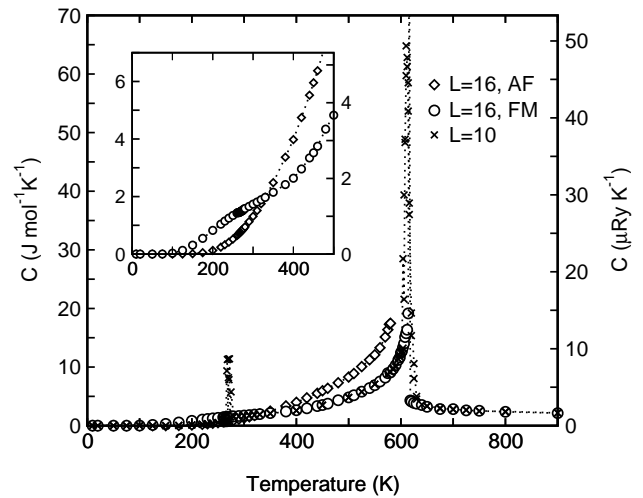}\hfill
  \end{center}  
  \caption{Specific heat as a function of temperature for the
    FM and the AF phases ($L=16$) and for simulations on smaller lattices 
    using the global MC step ($L=10$).
    Inset: The specific heat of the AF and FM phase between $0$ and
    $500\,$K.
    Clearly visible is a Schottky-type enhancement below $300\,$K in the FM
    phase, which is responsible for the metamagnetic transition.
    The sharp peak around $270\,$K
    corresponds to the metamagnetic transition, which is observed
    in the small system due to the global MC step.
}
\label{fig:C}
\end{figure}

A close look at the specific heat (Fig.~\ref{fig:C}) helps to
explain the occurrence of a metamagnetic transition. At first sight
it seems as if the specific heats of FM and AF phases do
not differ very much below $350\,$K. Above this temperature the
specific heat of the AF
phase increases more rapidly with temperature until around $570\,$K
the overheated AF phase is not stable anymore in the
simulations and transforms to the FM phase.
However, the inset shows that starting around $100\,$K
the specific heat of the FM phase is enhanced compared to the AF heat.
We explain this enhancement by a Schottky-type anomaly
which adds up to the excitations from spin flips.
Schottky anomalies are observed in
systems with two levels separated by a small energy barrier. 
In fact a crossover from magnetic to nonmagnetic Rh atoms
is conform with this picture. In the FM phase the Rh
atoms have a moment being ferromagnetically aligned to the Fe
moments, because the loss of the moment would correspond to an energy loss
of eight times the exchange constant. This amount is diminished by the
energy gain due to the $D_{\rm Rh}$ term, which
is smaller than $8\times J^{\rm nn}_{\rm FeRh}$ but larger than 
$4\times J^{\rm nn}_{\rm FeRh}$, since the ground state is otherwise not
antiferromagnetic. Increasing fluctuations of the
magnetization of the Fe sublattice cause
this energy difference to decrease and finally lead to a breakdown of the
average Rh moment at $T_{\rm C}$.
In the AF phase magnetic Rh atoms can be excited
at the expense of the energy $D_{\rm Rh}$. There is, however,
no gain in energy due to the exchange interaction, since the
contributions from the AF Fe atoms cancel
at the Rh site. This corresponds to a much larger energy difference
between magnetic and nonmagnetic Rh states compared to the FM phase and
does not lead to an appreciable contribution to the specific heat.  

\begin{figure}
  \begin{center}  
    \epsfig{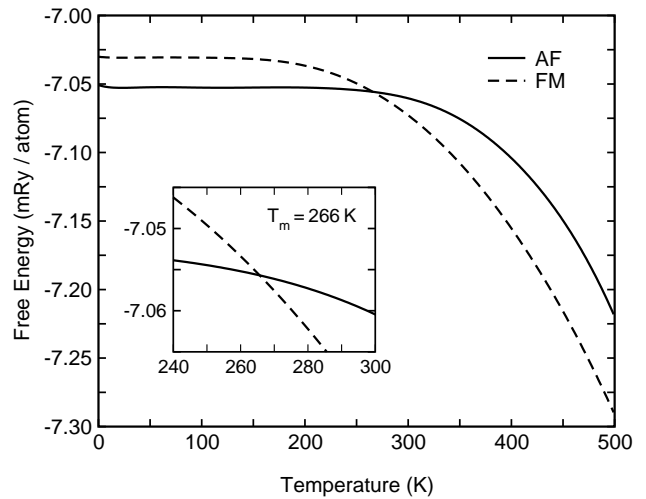}\hfill
  \end{center}  
  \caption{Free energy of the FM and AF phases obtained by integrating
    $C/T$, with the transition point at
    $T_{\rm m}=266\,$K.
    }
\label{fig:F}
\end{figure}

This view is further supported by a comparison of the free energy
of both phases shown in Fig.~\ref{fig:F}: The curves
intersect at $T\approx 266\,$K, which is in excellent agreement
with the simulated $T_{\rm m}=268\,$K. This accuracy could be achieved,
because the data for the specific heat obtained from the simulations have
only little spread and the fitted curves interpolate the data points
perfectly. From the difference between the
internal energies of both phases at the transition point, the entropy
jump at $T_{\rm m}$ is determined to $\Delta S(T_{\rm m})=5.2\,{\rm
J\,Kg^{-1}K^{-1}}$, which is only $30$ to $45\,\%$ of the
experimental values (but of the right order of magnitude).
But, one has to bear in mind that with the choice of an
Ising model we neglected the possibility of non-collinear
moments.
For the Rh moments this can only occur in the FM phase, and would
therefore contribute to $\Delta S(T_{\rm m})$. A second point is that the
weight of the $S_i=0$ state is the same as for each magnetic state
$S_i=\pm 1$. This is a natural choice for the spin-1 Ising model; but
for the real system this is somehow arbitrary, because we have no
information about the electronic origin of both Rh states.

\section{An extended model}\label{sec:extended}
Since the simple spin Hamiltonian~(\ref{eq:simple}) can
reproduce a metamagnetic transition as observed in $\alpha$-FeRh, it
remains an interesting question whether other outstanding properties
of this alloy, as the large volume increase at $T_{\rm m}$,
can also be explained.
Furthermore, it has still to be proven whether our spin analogy
is a good approximation to the {\em ab initio} results in the sense that it has
comparable low temperature properties.
In order to check this we have to extend (\ref{eq:simple}) for a
description of elastic and magneto-volume properties:
\begin{eqnarray}
      H &  = & -\,\sum_{i}^{~}\:D_i\:S_i^2 -
      \!\!\!\sum_{<{\rm nn,nnn}>}\!\!\!J_{ik}(r_{ik})\:S_i\:S_k
      \nonumber \\  
      & & + \sum_{<{\rm nn}>}\!\!V_{\rm nn}(r_{ik})
             + \!\!\sum_{<{\rm nnn}>}\!\!V_{\rm nnn}(r_{ik})\:.
             \label{eq:ext}
\end{eqnarray}
For $V(r_{ik})$ we use simple pair potentials of the
Lennard-Jones type:
\begin{equation}
 V(r_{ik}) = 4\:\epsilon\left[
        \left(\frac{\sigma}{r_{ik}}\right)^{12}-
        \left(\frac{\sigma}{r_{ik}}\right)^6\right] \:.
\end{equation}
Since in general the lattice structure of a Lennard-Jones system is closely
packed, two different pair potentials for nearest and next-nearest
neighbors have to be used in order to stabilize the bcc structure. The
potentials, however, do neither distinguish between different atom
types nor between the different spin states, as has been done in
previous simulations of related materials like Fe-Ni Invar
or Y(Mn$_x$Al$_{1-x}$)$_2$.\cite{cn:Gruner98,cn:Gruner02}
The use of Lennard-Jones
potentials is far from being optimum for metals, but has
numerical advantages that enable us to speed up the calculations
substantially. It is then
sufficient to choose the parameters so that basic elastic
properties like low temperature lattice constant, bulk modulus or
thermal expansion are reproduced.

Another change compared to (\ref{eq:simple}) is that the exchange
parameter is now taken to be a function
of interatomic distance. For simplicity, we assume a
linear distance dependence:
\begin{equation}
        J_{ik}(r_{ik}) = J_{ik}+
        \frac{\partial J_{ik}}{\partial r}\:r_{ik} \:. \label{eq:J}
\end{equation}
The values for $J_{ik}$ which we take in the extended calculations are
roughly the same as in Table~\ref{tab:param_simple}.
The relation between the derivatives of
$J^{\rm nn}_{\rm FeRh}$ and $J^{\rm nnn}_{\rm FeFe}$
with respect to the interatomic distance was
determined by the relation of the pressure derivatives of the Curie
and N\'eel temperatures, respectively, which have been obtained from
the experimental phase diagram by assuming a
linear dependence between $\partial J_{ik}/\partial r$ and
$\partial T_{\rm C,N}/\partial p$. The absolute values have been
adapted to reproduce the volume jump at $T_{\rm m}$.
An exchange interaction of this form has also been used to describe
magneto-volume effects in Fe-Ni
Invar.\cite{cn:Grossmann96,cn:Rancourt96,cn:Lagarec00}
In this case, the derivative of the exchange constant
$\partial J_{ik}/\partial r$
was a factor of $2$ to $20$ larger than in the present work.
The values of parameters
are summarized in Table~\ref{tab:param_ext}.
\begin{table}
\begin{ruledtabular}\begin{tabular}{rrrrr}
$D_{\rm Fe}$ & $D_{\rm Rh}$ &
$J^{\rm nn}_{\rm FeRh}({2.6\:{\rm \AA}})$ &
$J^{\rm nnn}_{\rm FeFe}({3.0\:{\rm \AA}})$ & 
$J^{\rm nnn}_{\rm RhRh}({3.0\:{\rm \AA}})$ \\ \hline 
$\gg\!k_{\rm B}\,T$ & $-11.1$ &
$2.10$ & $-1.04$ & 
$0$ \\[1ex]
 & & $\partial J^{\rm nn}_{\rm FeRh}/\partial r$
 &  $\partial J^{\rm nnn}_{\rm FeFe}/\partial r$
 &  $\partial J^{\rm nnn}_{\rm RhRh}/\partial r$ \\ \cline{3-5}
 & &  $1.97$ & $1.58$ & $0$ \\[1ex]
 & $\epsilon_{\rm nn}$ & $\sigma_{\rm nn}$ &
   $\epsilon_{\rm nnn}$ & $\sigma_{\rm nnn}$ \\ \cline{2-5}
 & $25.23$ & $2.32$ & $25.23$ & $2.67$
   \\
\end{tabular}\end{ruledtabular}
  \caption{Magnetic and elastic parameters for the extended spin model
      (energies in mRy, distances in $\rm\AA$).}
    \label{tab:param_ext}
\end{table}

\subsection{Details of computation}
For the evaluation of Hamiltonian (\ref{eq:ext}) we use a textbook
isothermal-isobaric Monte Carlo method (e.\,g.\ Ref.~\onlinecite{cn:Allen87})
consisting of alternating spin and position updates for each atom
and a global volume update step after finishing each lattice sweep.
This algorithm has been used by the the authors in previous
calculations and is explained in detail in the corresponding
references.\cite{cn:Gruner98,cn:Gruner98b,cn:Gruner02}
In order to simulate the metamagnetic transition, an additional global
spin update step has to be introduced.
We use the same algorithm as described in the last section for the spin system 
in connection with a simultaneous volume adaption.
Since the latter reduces the acceptance probability considerably,
the global update scheme has to be repeated several thousands of
times.
It is not practicable to use a new spin configuration for
each trial step, because the evaluation of the energy is
comparatively time consuming. On the other hand, the new energy
after solely rescaling the volume can be calculated very quickly,
since due to the use of Lennard-Jones potentials the energy can be written
as a function of integral powers of the lattice parameter.
Therefore, we choose a trial spin configuration as
described before and compute the new energy. Then we attempt
a previously fixed number $N$ of Metropolis steps, each with a newly
chosen volume. If one step is accepted, we continue with the original spin
configuration as trial system (and so on) until $N$ steps have been
made. Since the number of trial steps $N$ has been previously fixed,
{\em detailed balance} is still valid.

We have performed simulations with system sizes of $L=6$ to $12$.
A direct metamagnetic transition could only be seen for system sizes up to
$L=8$. The simulation time ranged from $120\,000$
up to $1\,000\,000$ lattice sweeps around $T_{\rm m}$ with values from
$1000$ to $10\,000$ for $N$. As before we
estimated Gibbs' free energy for zero pressure by integrating the
specific heat according to Eq.~(\ref{eq:F}).

\subsection{Computational results}
\begin{figure}
  \begin{center}  
    \epsfig{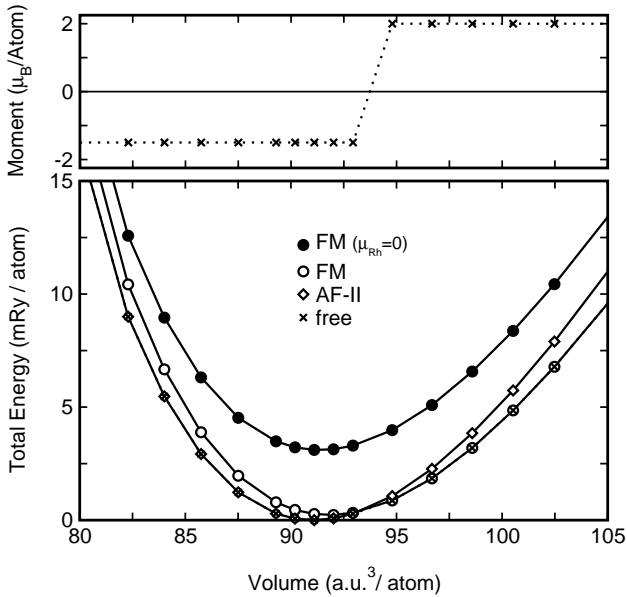}\hfill
  \end{center}  
  \caption{Internal energy and moment as a function of the atomic volume
    obtained for Hamiltonian (\ref{eq:ext}) at $T=4\,$K.
    The simulations have been performed for fixed spin structures as
    well as
    for a freely relaxed spin system.
    }
  \label{fig:T4}
\end{figure}
For a comparison of our model properties with the results
of ab initio calculations, we calculated in
isochoric simulations the energy as a function of the volume at low
temperatures for different fixed spin structures  (Fig.~\ref{fig:T4}).
Here, we cooled a system of size $L=16$ exponentially down from
$100\,$K down to $4\,$K. We find good agreement with the results of
Fig.~\ref{fig:ASW} in the sense that the order of magnetic phases
is similar.
Although the {\em ab initio} total energy differences are somewhat larger,
this indicates that our model Hamiltonian (\ref{eq:ext})
is a qualitatively correct description of the mechanisms leading to
a metamagnetic transition in FeRh.

\begin{figure}
  \begin{center}  
    \epsfig{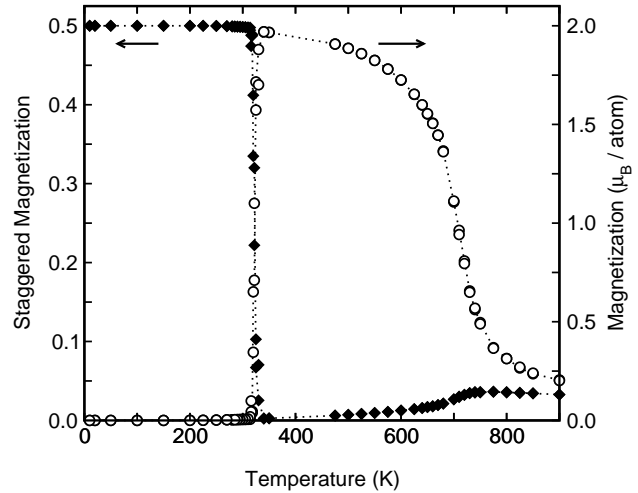}\hfill
  \end{center}  
  \caption{AF order parameter (diamonds) and magnetization 
    (circles) as a function
    of temperature as obtained for the extended model Hamiltonian
    (\ref{eq:ext}). The system size is $L=8$.
    }
  \label{fig:P00M}
\end{figure}
\begin{figure}
  \begin{center}  
    \epsfig{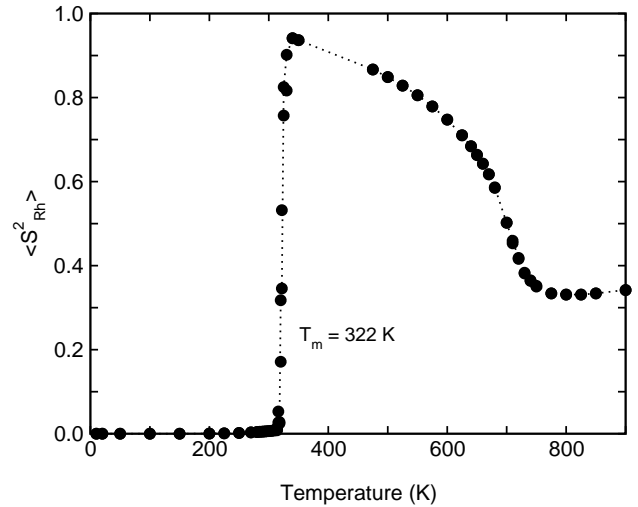}\hfill
  \end{center}  
  \caption{Mean magnetic moment of the Rh atoms for the extended
    model ($L=8$).
    }
  \label{fig:P00S2}
\end{figure}
As in the simple
model without volume-dependent terms,
we find an abrupt increase of the
magnetization in combination with a discontinuous decrease for the staggered
magnetization with increasing temperature (Fig.\ \ref{fig:P00M}).
Consequently the mean moment at the Rh sites
(Fig.~\ref{fig:P00S2}) also raises sharply around
the metamagnetic transition temperature of
$T_{\rm m}=322\,$K. Around the Curie temperature
$T=720\,$K the magnetization decreases more
smoothly than for the simple model. This may be due to the enhancement
of the effective exchange parameter given by Eq.~(\ref{eq:J}) which is
caused by the lattice expansion.

\begin{figure}
  \begin{center}  
    \epsfig{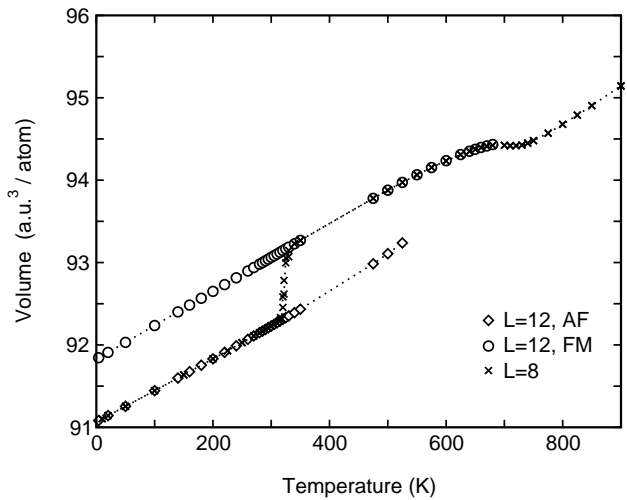}\hfill
  \end{center}  
  \caption{Volume per atom plotted over temperature as obtained for
     the AF and FM phases ($L=12$) and by our simulations with the global
     update scheme ($L=8$).
    }
  \label{fig:P00V}
\end{figure}
As expected from the low temperature calculations, the volume 
of the AF phase is smaller than the volume of the FM phase throughout the
stability range (Fig.~\ref{fig:P00V}).
For the freely fluctuating system a volume jump of $0.8\,\%$ occurs at
$T_{\rm m}$. Below $T_{\rm C}$ the volume expansion is reduced in the FM phase,
which is in qualitative agreement with 
experiment\cite{cn:Zakharov64TR,cn:Aigarabel96}.

\begin{figure}
  \begin{center}  
    \epsfig{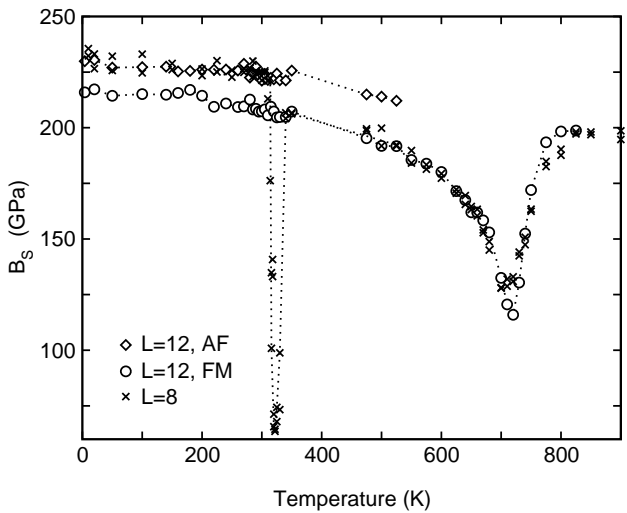}\hfill
  \end{center}  
  \caption{Adiabatic bulk modulus as a function of temperature for
     the AF and FM phases ($L=12$) and with the global
     update scheme ($L=8$). The latter data, which show the 
     metamagnetic transition, reveal a very large softening of the
     lattice around $T_{\rm m}$.}
  \label{fig:P00BS}
\end{figure}
The bulk modulus $B_{\rm S}$ of the AF phase is about $6\,\%$ larger
than the FM bulk modulus throughout the stability range (Fig.~\ref{fig:P00BS}).
Around $T_{\rm m}$ and $T_{\rm C}$ we find a considerable weakening of the
material which is caused by the magneto-volume anomalies.
The values for $B_{\rm S}$ at low temperatures can be estimated more
accurately by using a polynomial fit to the $E$-$V$ curves in
Fig.~\ref{fig:T4}. We find $B_{\rm S}=231\,$GPa for the AF phase and
$B_{\rm S}=217\,$GPa for the FM phase. These values are in good agreement with
results of isobaric calculations shown in Fig.~\ref{fig:P00BS}.
Compared with the results of {\em ab initio} calculations
(Table~\ref{tab:ab_initio}) the absolute values are too
large, whereas these calculations do not give a unanimous prediction for
the sign and the magnitude of the difference $\Delta B_{\rm S}$
(with respect to the magnetic structures).
Experimental measurements\cite{cn:Zakharov64TR} of Young modulus
suggest that the bulk modulus of the AF phase should in fact be lower
than in the FM phase. The modulus of the AF phase, however, has
only been estimated in the vicinity of $T_{\rm m}$, where a weakening of
the material (as a precursor of the transition) is present (as
in our simulations).

\begin{figure}
  \begin{center}  
    \epsfig{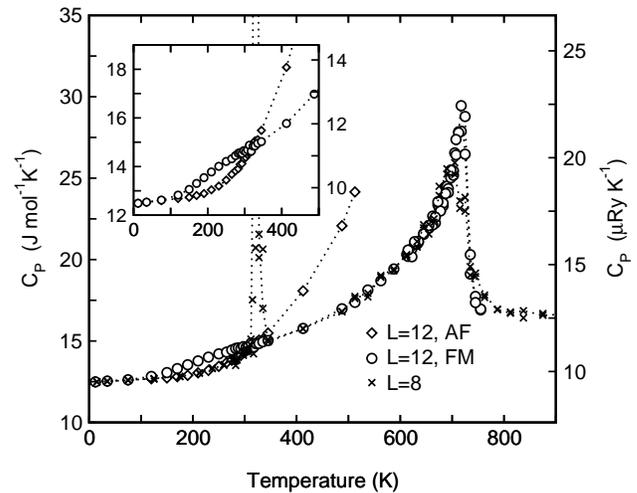}\hfill
  \end{center}  
  \caption{Specific heat for the extended model
    as a function of temperature for
    FM and the AF phases ($L=12$) and for simulations on smaller lattices 
    using the global MC step ($L=8$). The relationship between 
    the specific heats
    of AF and FM phases below $400\,$K is shown in the inset.
    }
  \label{fig:P00Cp}
\end{figure}
The specific heats obtained for the extended model 
(Fig.~\ref{fig:P00Cp}) resemble the findings for the spin-only
Hamiltonian~(\ref{eq:simple}), except that now additional contributions
due to atomic displacements are included.
This shows that the
proposed explanation of the metamagnetic transition is still valid.
Since the Hamiltonian (\ref{eq:ext}) does not contain any kinetic terms,
the low temperature value of the specific heat is only half of the
Dulong-Petit limit of $3\,R$, where $R$ is the kinetic gas constant.

\subsection{Estimation of Gibbs free energy}
\begin{figure}
  \begin{center}  
    \epsfig{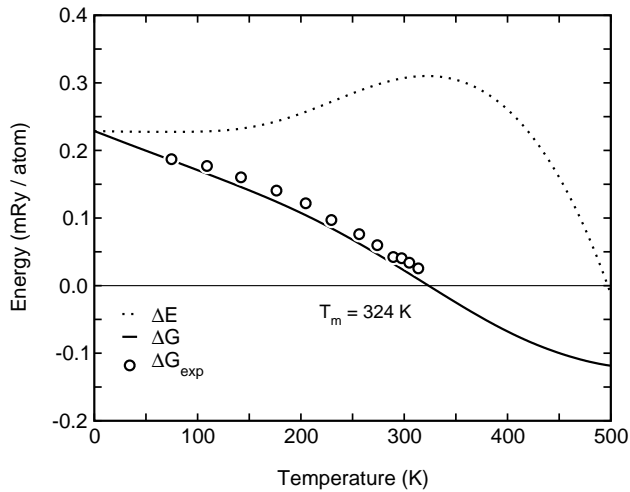}\hfill
  \end{center}  
  \caption{Differences of internal energies and Gibbs free
    energies for FM and AF phases obtained by integrating
    $C_{\rm p}/T$. The transition point is
    $T_{\rm m}=324\,$K. The experimental data have been obtained by
    integrating over the hysteresis loop for different
    temperatures.\cite{cn:Ponomarev73TR}
    }
  \label{fig:P00F}
\end{figure}
The calculation of Gibbs free energy $G$ for the model
with classical motion (\ref{eq:ext}) is more
complicated than in the spin-only case (\ref{eq:simple}).
First of all absolute values of $G$ cannot be given, since the
specific heat at zero temperature is finite and the resulting entropy
would diverge. But since the limit for $T$\,$\to$\,$0$ of $C_{\rm p}$ is
$1.5\,R$ and hence the same for all magnetic structures,
differences of free energies can be computed.
In contrast to the previous section, the entropy contribution at zero
temperature $\Delta S(0)$ must be considered.
As before a contribution from the magnetic system can be neglected,
while we have to account for the differences in the elastic
properties of FM and the AF phases. Since we only need the entropy
at zero temperature, we choose an ensemble of
harmonic oscillators as an approximation for our spatial degrees of
freedom, neglecting the anharmonicity of the potentials and the
coupling of the oscillators.
Differentiating the free energy obtained from the logarithm of the
partition function leads then to a simple expression for the entropy
difference,
\begin{equation}
\Delta S=\frac{3}{2} N k_{\rm B}\ln\left(k_{\rm AF}/k_{\rm FM}\right)\:,
\end{equation}
where $k_{\rm AF}$ and $k_{\rm FM}$ are the force constants of the harmonic
potentials which are estimated from the curvature of the
ground state energy (versus lattice parameter) curve.
For the entropy difference we obtain then:
$\Delta S(0)=9.17\,{\rm J\,kg}^{-1}{\rm K}^{-1}$. Taking this into
account, we achieve again a rather good value for $T_{\rm m}$.
From the differences of internal energies at $T_{\rm m}$ we obtain
for the entropy jump at the transition
$\Delta S(T_{\rm m})=15.9\,{\rm J\,kg}^{-1}{\rm K}^{-1}$,
which is within the range of experimental results. 
Comparison of the calculated free energy
with experimental values for $\Delta G$ shows excellent agreement.
The experimental values have been obtained by a graphical integration 
of the magnetic field expressed as a function of the measured
magnetization.\cite{cn:Ponomarev73TR}
Extrapolation of experimental values to zero temperature
shows that the energy difference $\Delta E(0)$ between the AF and 
FM phases is much better described by our model parameters than 
by {\em ab initio} results, since the latter show that $\Delta E$ is
one order of magnitude too large. This discrepancy has already been noticed by
Moruzzi and Marcus\cite{cn:Moruzzi92b} who relate this to the omission
of zero point energy corrections in their total energy calculations.

\section{Conclusions}\label{sec:conclusions}
We propose on the basis of new {\em ab initio} results a
mechanism for the metamagnetic transition in FeRh at finite temperatures. In
contrast to previous explanations, our model does not rely on a large
difference between the low temperature specific heat constants
of both phases. These are expected to be sensitive to external
influences as both experimental measurements and band structure calculations
suggest, so that it seems implausible that a constant contribution of
the given magnitude might survive up to room temperature.
Instead, we propose that the existence of two magnetic states of
Rh atoms connected with competing FM
Fe-Rh and AF Fe-Fe exchange interactions are at the origin
of the metamagnetic transition. The magneto-volume effects can
simply be explained on the basis of distance dependent exchange parameters.
The applicability of this mechanism to the Fe-Rh problem has been
verified by Monte Carlo model calculations, showing that a
metamagnetic transition of the desired kind does in fact occur
and, by extending the model, magneto-volume effects and other
experimental properties can be sufficiently well described.

As we have pointed out, our explanation is in agreement
with existing experimental data. However, a 
further check of our model
would be a comparison with (non-existing)
specific heat data from above $T_{\rm C}$
down to very low temperatures for both, the AF and FM phases.
From this one could then estimate the magnetic contribution by subtracting the
lattice part within the Debye approximation, the electronic part and
the contribution by the anharmonicity of the
potentials.\cite{cn:Bendick78,cn:Rellinghaus95}
So far, the specific heat has only been determined in the range 
from $100\,$K to $500\,$K
and only for the nearly stoichiometric Rh-rich
alloy.\cite{cn:Richardson73,cn:Annaorazov92} Systematic
measurements on the Fe-rich side with a FM ground state and
measurements under pressure, suppressing the metamagnetic transition
would yield information whether a Schottky-type excitation plays an
important role, which should show up around $200\,$K in the
magnetic contribution to the specific heat of ferromagnetic samples.

Monte Carlo simulations with applied pressure are left
for future work, since with increasing pressure and hence increasing
$T_{\rm m}$ a reliable estimation of the transition temperature is rather
difficult and requires an
improvement of the global MC step. First tests,
however, showed that the location of the phase boundaries under
pressure is in sufficient agreement with experimental data for
pressures below $20\,$kbar and above $40\,$kbar, where the metamagnetic
transition is completely suppressed. The tricritical point, if one
exists, should
be located somewhere between $20\,$kbar and $40\,$kbar for the
parameters used here. 

\begin{acknowledgments}
This work has been supported by the DFG (Deutsche
Forschungsgemeinschaft) through the SFB (Sonderforschungsbereich) 445
and the Graduate College {\em Structure and Dynamics of
  Heterogeneous Systems}.
\end{acknowledgments}


\end{document}